\documentclass[usegraphicx,usenatbib,useAMS]{mn2e}
\title{On model selection forecasting, Dark Energy and modified gravity }
\author[Heavens et al]{A. F. Heavens$^1$, T.D. Kitching$^2$ and L. Verde$^3$\\
$^1$ SUPA\footnote{Scottish Universities Physics Alliance},
Institute for Astronomy, University of Edinburgh, Royal
Observatory, Blackford Hill, Edinburgh, EH9S 3HJ, UK,\\
$^2$ University of Oxford, Denys Wilkinson Building, Department of Physics,
Wilkinson Building, Keble Road, Oxford OX1 3RH, UK,\\
$^3$ Department of Physics, University of Pennsylvania, 209 South
33rd St, Philadelphia, PA 19104-6396, USA}

\begin{document}

\maketitle

\begin{abstract}
The Fisher matrix approach (\cite{Fisher}) allows one to calculate
in advance how well a given experiment will be able to estimate
model parameters, and has been an invaluable tool in experimental
design. In the same spirit, we present here a method to predict how
well a given experiment can distinguish between different models,
regardless of their parameters. From a Bayesian viewpoint, this
involves computation of the Bayesian evidence. In this paper, we
generalise the Fisher matrix approach from the context of parameter
fitting to that of model testing, and show how the expected evidence
can be computed under the same simplifying assumption of a gaussian
likelihood as the Fisher matrix approach for parameter estimation.
With this `Laplace approximation' all that is needed to compute the
expected evidence is the Fisher matrix itself. We illustrate the
method with a study of how well upcoming and planned experiments
should perform at distinguishing between Dark Energy models and
modified gravity theories. In particular we consider the combination
of 3D weak lensing, for which planned and proposed wide-field
multi-band imaging surveys will provide suitable data, and probes of
the expansion history of the Universe, such as proposed supernova
and baryonic acoustic oscillations surveys.  We find that proposed
large-scale weak lensing surveys from space should be able readily
to distinguish General Relativity from modified gravity models.
\end{abstract}


\section{Introduction}

While the goal of parameter estimation is to determine the best fit
values (and the errors) of a set of parameters within a model, model
selection seeks to distinguish between different models, which in
general will have different sets of parameters. Model selection has
received attention in cosmology only relatively recently (starting
with \cite{Jaffe96}). Only in recent years have cosmological data
had enough statistical power to address the problem, although
elsewhere in astronomy model selection has been applied for some
time (e.g. \cite{Lucy71}).

For parameter estimation, the Fisher matrix approach (\cite{Fisher})
has been invaluable in experimental design. It allows one to
forecast how well a given experiment will be able to estimate model
parameters. In the same spirit, one may want to predict how well a
given experiment can distinguish between different models.

Of particular interest is the case of {\em nested} models, where the
more complicated model has additional parameters, in addition to
those in the simpler model. The simpler model may be interpreted as
a particular case for the more complex model where the additional
parameters are kept fixed at some fiducial values. The additional
parameters may be an indication of new physics, thus the question
one may ask is: ``would the experiment provide data with enough
statistical power to require additional parameters and therefore to
signal the presence of new physics if the new physics is actually
the true underlying model?''

Examples of this type of questions are: ``does the primordial power
spectrum deviate from scale invariance?'', ``do the data require a
running of the primordial power spectrum spectral index?'', ``do
observations require a dark energy that deviates from a cosmological
constant with equation of state parameter $w=-1$?'' etc. (e.g.
\cite{LMPW, BLH06, Mukherjeeetal06,  PML06, Saini, BLH07, Trotta07}
and references therein).

In this paper we present a general statistical method for rapidly
calculating in advance how well a given experimental design can be
expected to distinguish between competing (and in particular nested)
theoretical models for the data, under the Laplace approximation.

We then present an application of the method to the cosmological
context: we forecast how three proposed weak lensing experiments may
distinguish between Dark Energy (e.g. \cite{PeeblesRatra,Wetterlich}
and modified gravity models (e.g. \cite{DGP} (DGP)).

The standard cosmological model is extremely successful: with only 6
parameters it fits a host of observations and provides a description
of the Universe from z=1100 to present day.  This finding is in good
agreement with what can be expected of current data, based on
Bayesian complexity theory ((\cite{KTP}).  The parameters of the
model are tightly constrained, many at the percent level, and there
is a considerable weight of evidence in favour of a substantial
contribution of Dark Energy. Alternative explanations for the
apparent Dark Energy are a Cosmological Constant, a slowly rolling
scalar field or a fluid component, whose effects on the expansion
history can be described by an equation of state parameter which may
evolve in time, or a large-scale modification to General Relativity.

Each of these can be described as a different model. An interesting
question is which of these possibilities is favoured by the data. In
contrast to {\em parameter estimation}, this is an issue of {\em
model selection}, which has been the subject of recent attention in
cosmology (e.g.
\cite{Hobson,Saini,MPL06,LMPW,Szydlowski06a,Szydlowski06b,Pahud06,Pahud07,Serra07}
and references therein).  In particular, \cite{Mukherjeeetal06, Trotta06}
compare evolving dark energy models with a cosmological constant.

Dark Energy has potentially measurable effects through its effect on
both the expansion history of the Universe, and through the growth
rate of perturbations. Some methods, such as study of the luminosity
distance of Type Ia supernovae (SN; e.g. \cite{Riess}), baryonic
acoustic oscillations (BAO; e.g. \cite{EisensteinHu98}) or geometric
weak lensing methods (e.g. \cite{TKBH07}), probe only the expansion
history, whereas others such as 3D cosmic shear weak lensing or
cluster counts can probe both. Combinations of various probes
promise very accurate determinations of the equation of state (e.g.
\cite{Heavens06}).

For methods based on probing the expansion history alone, there is a
difficulty in that the same expansion history of a universe where
the law of gravity has been modified on large scales can also be
obtained in a universe with standard General Relativity but a dark
energy component with a suitable equation of state parameter $w(z)$.
In general however the growth history of cosmological structures
will be different in the two cases (e.g. \cite{HutererLinder06}, but
see \cite{Kunz}).

Therefore in principle there are advantages in using methods which
also probe the growth rate, such as 3D weak lensing.  It becomes a
very interesting question to ask whether such methods could
distinguish between the Dark Energy and modified gravity scenarios.

This question may be answered in a Bayesian context by considering
the Bayesian Factor, which is the ratio of the Bayesian Evidences,
i.e. the ratio of probabilities of the data given the two models.
The evidence ratio may be generalised to a genuine posterior
probability of the models by multiplying by a ratio of priors of the
models. The evidence involves an integration of the likelihood,
multiplied by priors, over the parameter space of each model, and
this can be computationally expensive if the dimension of the
parameter space is large. By making a simplifying assumption in the
spirit of Fisher's analysis, one can compute the expected evidence
for a given experiment, in advance of taking any data, and forecast
the extent to which an experiment may be able to distinguish between
different models.  The Fisher matrix approach in parameter
estimation assumes that the expected behaviour of the likelihood $L$
near the maximum characterises the likelihood sufficiently well to
be used to estimate errors on the parameters (i.e. the Taylor
expansion of $\ln L$ to include second-order terms holds at least
until $\ln L$ drops by $\sim 1$).  The Fisher Matrix is defined by
\begin{equation}
F_{\alpha\beta} \equiv -\left\langle \frac{\partial^2 \ln
L}{\partial \theta_\alpha \theta_\beta}\right\rangle
\end{equation}
where $\theta_\alpha$ are the model parameters, and $\langle \ln
L\rangle \simeq \langle\ln L^0\rangle -
F_{\alpha\beta}\left[\theta_\alpha-\theta_\alpha^0\right]
\left[\theta_\beta-\theta_\beta^0\right]/2$, where $0$ indicates
peak values. In this paper, we compute the expected evidence by
assuming that we can ignore higher-order terms in the Taylor
expansion throughout. This allows a major simplification, called the
Laplace approximation, in that the expected evidence can be computed
directly from the Fisher Matrix at essentially no extra cost.  Of
course, one must make the caveat that the quadratic expansion of
$\ln L$ may be a poor approximation in some cases, but nevertheless
the computation of the expected evidence may be a useful first step.

The layout of this paper is that we present the formal computation
of the ratio of expected evidences (also called the Bayes factor) in
Section 2, and apply the method to a combination planned and
proposed data sets (3D weak lensing, microwave background
measurements, supernovae and baryonic acoustic oscillations probes)
in Section 3.

\section{Forecasting Evidence}

The aim here is to compute the Bayesian Evidence ratio for
two different models,  i.e. given a dataset arising from a true
model, we want to know the probability with which a second model can
be ruled out.

The spirit of this is very similar to the Fisher Matrix approach for
parameter estimation, where one computes the expected likelihood as
a function of parameters, in the absence of data.

We make the approximation that the expected likelihood is everywhere
accurately described by a multivariate Gaussian, with a curvature
given by a Taylor expansion at the expected peak. Thus it should be
regarded as a first step; more sophisticated simulation techniques
may be necessary if this assumption is not adequate.  A special case
of this was recently presented by \cite{Trotta07}.  However, it is
worth pointing out that it is routine to compute expected marginal
errors on parameters using the inverse of the Fisher matrix.  This
is equivalent to assuming the Laplace approximation, and doing only
one fewer integration than we propose here.  In some cases, the
approximation works very well (see e.g. \cite{LISA}), but there are
certainly some examples where the approximation is not particularly
accurate (e.g. \cite{Wang}).  For the {\em Planck} microwave
background experiment we find reasonable agreement between Fisher and
Monte Carlo Markov Chain errors (to within about 30\%), if the same
likelihood calculations are used for both.  Note that for this paper, we follow
\cite{Hu2002} for the likelihood, which is accurate for the
noise-dominated regime. For Planck, the errors will therefore be
rather conservative.

The other approximations we make are that the priors on the
parameters are uniform, but this could be relaxed to Gaussians if
desired. We also assume that the priors on the two models are the
same; this could easily be relaxed as it just adjusts the
normalisation.

\subsection{Models and Notation}

We denote two competing models by $M$ and $M'$.  We assume that $M'$
is a simpler model, which has fewer ($n'<n$) parameters in it.  We
further assume that it is {\em nested} in Model $M'$, i.e. the $n'$
parameters of model $M'$ are common to $M$, which has $p\equiv n-n'$
extra parameters in it. These parameters are fixed to fiducial values in $M'$.

We denote by $D$ the data vector, and by $\theta$ and $\theta'$ the
parameter vectors (of length $n$ and $n'$).

The posterior probability of each model comes from Bayes' theorem:
\begin{equation}
p(M|D) = \frac{p(D|M)p(M)}{p(D)}
\end{equation}
and similarly for $M'$.  By marginalisation $p(D|M)$, known as the
{\em Evidence}, is
\begin{equation}
p(D|M) = \int d\theta\,p(D|\theta,M)p(\theta|M),
\end{equation}
which should be interpreted as a multidimensional integration. Hence
the posterior relative probabilities of the two models, regardless
of what their parameters are, is
\begin{equation}
\frac{p(M'|D)}{p(M|D)}=\frac{p(M')}{p(M)}\frac{\int
d\theta'\,p(D|\theta',M')p(\theta'|M')}{\int
d\theta\,p(D|\theta,M)p(\theta|M)}.
\end{equation}
With non-committal priors on the models, $p(M')=p(M)$, this ratio
simplifies to the ratio of evidences, called the {\em Bayes Factor},
\begin{equation}
B \equiv \frac{\int d\theta'\,p(D|\theta',M')p(\theta'|M')}{\int
d\theta\,p(D|\theta,M)p(\theta|M)}.
\end{equation}
Note that the more complicated model $M$ will inevitably lead to a
higher likelihood (or at least as high), but the evidence will
favour the simpler model if the fit is nearly as good, through the
smaller prior volume.

We assume uniform (and hence separable) priors in each parameter,
over ranges $\Delta\theta$ (or $\Delta\theta'$).  Hence
$p(\theta|M)=(\Delta\theta_1\ldots \Delta\theta_n)^{-1}$ and
\begin{equation}
B = \frac{\int d\theta'\,p(D|\theta',M')}{\int
d\theta\,p(D|\theta,M)}\,\frac{\Delta\theta_1\ldots\Delta\theta_n}{\Delta\theta'_1\ldots\Delta\theta'_{n'}}.
\label{Bnew}
\end{equation}
Note that if the prior ranges are not large enough to contain
essentially all the likelihood, then the position of the boundaries
would influence the Bayes factor.  In what follows, we will assume
the prior range is large enough to encompass all the likelihood.

In the nested case, the ratio of prior hypervolumes simplifies to
\begin{equation}
\frac{\Delta\theta_1\ldots\Delta\theta_n}{\Delta\theta'_1\ldots\Delta\theta'_{n'}}=\Delta\theta_{n'+1}\ldots
\Delta\theta_{n'+p},
\end{equation}
where $p\equiv n-n'$ is the number of extra parameters in the more
complicated model.

The Bayes factor in equation (\ref{Bnew}) still depends on the specific dataset $D$.  For future experiments, we do not yet have the data, so we compute the expectation value of the Bayes factor, given the statistical properties of $D$.  The expectation is computed over the distribution of $D$ for the correct model (assumed here to be $M$).  To do this, we make two further approximations: first we note that $B$ is a ratio, and we approximate $\langle B\rangle$ by the ratio of the expected values, rather than the expectation value of the ratio.  This should be a good approximation if the evidences are sharply peaked.  

We also make the Laplace approximation, that the expected likelihoods are given by multivariate Gaussians.  For example,
\begin{equation}
\langle p(D|\theta,M)\rangle  = L_0 \exp\left[-\frac{1}{2}(\theta-\theta^0)_\alpha
F_{\alpha\beta}(\theta-\theta^0)_\beta\right],
\end{equation}
which is centred on $\theta^0$, the correct parameters in $M$.

\begin{figure}
  \includegraphics[width=0.45\textwidth]{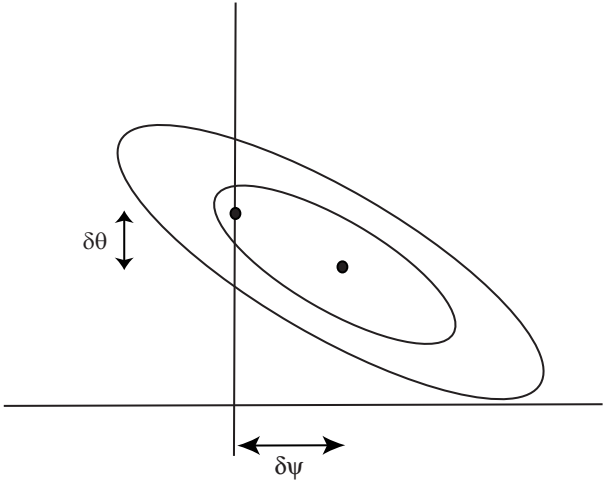}\\
  \caption{Illustrating how assumption of a wrong parameter value can influence the
  best-fitting value of other model parameters.  Ellipses represent iso-likelihood surfaces,
  and here in the simpler model, the parameter on the horizontal axis is assumed to take the
  value given by the vertical line. Filled circles show the true parameters in the more complicated model,
  and the best-fit parameters in the simpler model.}
  \label{offsetfig}
\end{figure}

A similar expression is assumed for $\langle p(D|\theta',M')\rangle$.  The Laplace approximation assumes that a Taylor expansion of the likelihood around the peak value to second order
can be extended throughout the parameter space. $F_{\alpha\beta}$ is
the Fisher matrix, given for Gaussian-distributed data by (see e.g.
\cite{TTH})
\begin{equation}
F_{\alpha\beta}=\frac{1}{2}{\rm
Tr}\left[C^{-1}C_{,\alpha}C^{-1}C_{,\beta}+C^{-1}(\mu_{,\beta}\mu^t_{,\alpha}+\mu_{,\alpha}\mu^t_{,\beta})\right].
\end{equation}
$C$ is the covariance matrix of the data, and $\mu$ its mean (no
noise). Commas indicate partial derivatives w.r.t. the parameters.
For the correct model $M$, the peak of the expected likelihood is located at the true parameters $\theta^0$.  Note, however, that for the incorrect model $M'$, the peak of the expected likelihood is not in general at the true parameters (see Fig. \ref{offsetfig} for an illustration of this).  This arises because the likelihood in the numerator of equation (\ref{Bnew}) is the probability of the dataset $D$ given incorrect model assumptions.

If we assume that the posterior probability densities are small at
the boundaries of the prior volume, then we can extend the
integrations to infinity, and the integration over the multivariate
Gaussians can be easily done.  This gives, for $M$, $(2\pi)^{n/2}
(\det{F})^{-1/2}$, so for nested models,
\begin{equation}
\langle B \rangle =
(2\pi)^{-p/2}\frac{\sqrt{\det{F}}}{\sqrt{\det{F'}}}\frac{L'_0}{L_0}\Delta\theta_{n'+1}\ldots
\Delta\theta_{n'+p}.
\end{equation}
An equivalent expression was obtained, using again the Laplace
approximation by \cite{Lazarides}. The point here is that with the
Laplace approximation, one can compute the $L'_0/L_0$ ratio from the
Fisher matrix.  To compute this ratio of likelihoods, we need to
take into account the fact that, if the true underlying model is
$M$, in $M'$ (the incorrect model), the maximum of the expected
likelihood will not in general be at the correct values of the
parameters (see Fig. \ref{offsetfig}). The $n'$ parameters shift
from their true values to compensate for the fact that, effectively,
the $p$ additional parameters are being kept fixed at incorrect
fiducial values. If in $M'$, the additional $p$ parameters are
assumed to be fixed at fiducial values which differ by
$\delta\psi_\alpha$ from their true values, the others are shifted
on average by an amount which is readily computed under the
assumption of the multivariate Gaussian likelihood (see e.g.
\cite{TKBH07}):
\begin{equation}
\delta\theta'_\alpha =
-(F'^{-1})_{\alpha\beta}G_{\beta\zeta}\delta\psi_\zeta \qquad
\alpha,\beta=1\ldots n', \zeta=1\ldots p \label{offset}
\end{equation} where
\begin{equation}
G_{\beta\zeta}=\frac{1}{2}{\rm
Tr}\left[C^{-1}C_{,\beta}C^{-1}C_{,\zeta}+C^{-1}(\mu_{,\zeta}\mu^t_{,\beta}+\mu_{,\beta}\mu^t_{,\zeta})\right],
\end{equation}
which we recognise as a subset of the Fisher matrix. For clarity, we
have given the additional parameters the symbol $\psi_\zeta; \
\zeta=1 \ldots p$ to distinguish them from the parameters in $M'$.

With these offsets in the maximum likelihood parameters in model
$M'$, the ratio of likelihoods is given by
\begin{equation}
L'_0 = L_0 \exp\left(-\frac{1}{2}\delta\theta_\alpha F_{\alpha\beta}
\delta\theta_\beta\right)
\end{equation}
where the offsets are given by $\delta\theta_\alpha =
\delta\theta'_\alpha$ for $\alpha\le n'$ (equation \ref{offset}),
and $\delta\theta_\alpha = \delta\psi_{\alpha-n'}$ for $\alpha>n'$.

The final expression for the expected Bayes factor is then
\begin{equation}
\langle B \rangle =
(2\pi)^{-p/2}\frac{\sqrt{\det{F}}}{\sqrt{\det{F'}}}\exp\left(-\frac{1}{2}\delta\theta_\alpha
F_{\alpha\beta}\delta\theta_\beta\right)\prod_{q=1}^p\Delta\theta_{n'+q}.
\label{Final}
\end{equation}
Note that $F$ and $F^{-1}$ are $n \times n$ matrices, $F'$ is $n'
\times n'$, and $G$ is an $n' \times p$ block of the full $n \times
n$ Fisher matrix $F$.  The expression we find is a specific example
of the Savage-Dickey ratio (\cite{Dickey}); here we explicitly use
the Laplace approximation to compute the offsets in the parameter
estimates which accompany the wrong choice of model.

Note that the `Occam's razor' term, see \cite{Saini} for example,
common in evidence calculations, is encapsulated in the
$(2\pi)^{-p/2}\frac{\sqrt{\det{F}}}{\sqrt{\det{F'}}}$ factor
multiplied by the prior product: models with more parameters are
penalised in favour of simpler models, unless the data demand
otherwise. Such terms should be treated with caution, as pointed out
by \cite{LinderMiquel07} simpler models do not always result in the
most physically realistic conclusions (but see \cite{Liddle07} for a
thorough discussion of the issues). In cases where the Laplace
approximation is not a good one, other techniques must be used, at
more computational expense (e.g. \cite{Trotta05,Trotta07,
Beltranetal05,Skilling04,MPL06,PML06,Mukherjeeetal06}).
Alternatively, a reparametrization of the parameter space can make
the likelihood closer to gaussian (see e.g. \cite{Kosowsky02} for
CMB).

According to \cite{Jeffreys61}, $1<\ln B<2.5$ is described as
`substantial' evidence in favour of a model, $2.5<\ln B< 5$ is
`strong', and $\ln B>5$ is `decisive'. These descriptions seem too
aggressive: $\ln B=1$ corresponds to a posterior probability for the
less-favoured model which is 0.37 of the favoured model
(\cite{Kass}). Other authors have introduced different terminology
(e.g. \cite{Trotta05}).

\section{Application: Dark Energy or Modified Gravity}

To apply these results to cosmological probes of Dark
Energy/modified gravity, we use the convenient Minimal Modified
Gravity parametrization introduced by \cite{Linder05} and expanded
by \cite{LinderCahn07} and \cite{HutererLinder06}, where the
beyond-Einstein perturbations are described by a growth factor
$\gamma$.  The growth rate of perturbations in the matter density
$\rho_m$, $\delta \equiv \delta \rho_m/\rho_m$, is accurately
parametrised as a function of scale factor $a(t)$ by
\begin{equation}
\frac{\delta}{a} \equiv g(a) =
\exp\left\{\int_0^a\,\frac{da'}{a'}\left[\Omega_m(a')^\gamma-1\right]\right\},
\end{equation}
where $\Omega_m(a)$ is the density parameter of the matter.  The
growth factor $\gamma\simeq 0.55$ for the standard General
Relativistic cosmological model, whereas for modified gravity
theories it deviates from this value.  For example, for the DGP
braneworld model (\cite{DGP}), $\gamma\simeq 0.68$
(\cite{LinderCahn07}), on scales much smaller than those where
cosmological acceleration is apparent. For this paper, we introduce
$\gamma$ as an additional parameter in Model $M$ - i.e. $M$
represents extensions beyond General Relativity, whereas $M'$
represents General Relativity with Dark Energy. \cite{SongHuSawicki}
show that the simplest flat DGP model demonstrates a difficulty
pointed out by \cite{IshakUpadyeSpergel}, i.e. inconsistent
parameter values are obtained from different datasets. Also the DGP
model may be in difficulties with the CMB peak and baryon
oscillations (\cite{Rydbeck}), as well as theoretically through the
existence of ghosts (\cite{Gorbunov}). Here we concentrate instead
on distinguishing modified gravity models from GR using a different
feature - that the growth factor is different from that of a
quintessence model with the same expansion history. We use the DGP
model as a specific example of a more general test of modified
gravity models.

We make the (conservative) assumption that modification of gravity
does not change the growth factor of perturbations on scales
comparable to the horizon, i.e. in all cases the Integrated Sachs
Wolfe (ISW) effect at large CMB angular scale is computed assuming
the perturbation growth is given by standard gravity. In addition
for the calculation of the CMB low $\ell$ multipoles we assume the
dark energy perturbations associated to a scalar field with the same
effective equation-of-state parameters as the modified gravity
model. This is a conservative assumption as in a modified gravity
model the ISW effect is expected to be different from standard
gravity and therefore CMB observations may have some extra
sensitivity to $\gamma$ which we ignore here.

The full set of parameters we explore in $M'$ is $\Omega_m,
\Omega_b, h, \sigma_8, n_s, \alpha_n, \tau, r, w_0, w_a$, being the
density parameters in matter and baryons, the Hubble constant (in
units of 100 km$^{-1}$Mpc$^{-1}$), the amplitude of fractional
density perturbations, the primordial scalar spectral index of
density fluctuations, and its running with wavenumber $k$, the
reionisation optical depth, the tensor-to-scalar ratio; finally,
there are two parameters characterising the expansion history of the
Universe, $w(a)= w_0 + w_a(1-a)$ (\cite{Chevallier}).  For Dark
Energy models, this is the equation of state parameter $p/(\rho
c^2)$ as a function of scale factor $a$.  However, it is used here
only as a means of parametrising of the expansion history, in terms
of an effective Dark Energy component - $w(a)$ is not necessarily
associated with a Dark Energy component. See
\cite{HutererLinder06,LinderCahn07} and \cite{Kunz}, For example, in
the DGP model the expansion history is described well by
$w_0=-0.78$, $w_a=0.32$.  The Fisher matrices are almost unchanged
if we take this as the fiducial model, so we present results for
$w_0=-1$, $w_a=0$.  $\gamma$ is an additional parameter in $M$ (set
fixed at 0.55 in $M'$), which parametrises the growth of structure
For completeness, we list here the other fiducial parameters:
$\Omega_m=0.27, \Omega_b=0.04, h=0.71, \sigma_8=0.8, n_s=1.0,
\alpha_n=0.0, \tau=0.09, r=0.01$.

Thus the question we want to address is the following: assuming that
the model of the Universe is a modified gravity model, is there an
experimental setup which can distinguish this model from a Dark
Energy model with the same expansion history? In this application we
initially take the parameters of the model to be the DGP ones.

The experiments we consider are the upcoming \emph{Planck} microwave
background survey (\cite{Planck}), including polarisation
information, three 3D weak lensing surveys and proposed SN and BAO
surveys.  Note that, as discussed above, we set the CMB constraint
on $\gamma$ to zero.  The constraint on $r$ and $\tau$ from the weak
lensing is similarly zero - these are assumed fixed in the
weak-lensing alone experiments; in the weak-lensing plus CMB, the
constraints on $r$ and $\tau$ come from the CMB.

We consider a number of 3D weak lensing surveys: firstly a survey
covering $5000$ square degrees to a median redshift of $z_{m}=0.8$
with a source density of $10$ galaxies per square arcminute, such as
might be achieved with the Dark Energy Survey (DES) (\cite{DES});
second a survey covering $30,000$ square degrees of the sky
(\cite{PanSTARRS}) to a median depth of $z_{m}=0.75$ with $5$
galaxies per square arcminute, as might be achieved with Pan-STARRS
(we consider only the single-telescope Pan-STARRS 1); third is
a survey of $35$ sources per square arcminute, $z_{m}=0.90$, and an
area of $20,000$ square degrees (next generation weak lensing
survey, WL$_{NG}$), as might be observed by a space-based survey
such as \emph{DUNE}, which is a candidate for the ESA Cosmic Vision
programme, or the Supernova Acceleration Probe (\emph{SNAP}), a
candidate of the NASA Joint Dark Energy Mission. Note that the
characteristics of the Large Synoptic Survey Telescope (LSST) data
set are not too dissimilar from these, so the reported numbers would
be very close to those for LSST.  For all surveys we assume flatness
and a redshift dependence of source density $n(z) \propto
z^2\exp\left[-(z/z*)^{1.5}\right]$, with $z*=1.4 z_{m}$ and use the
3D weak shear power spectrum analysis method of \cite{Heavens03} and
\cite{Heavens06}.  The modes are truncated at $k=1.5$Mpc$^{-1}$,
avoiding the highly nonlinear regime where uncertainties in the
power spectrum may lead to dominant systematic errors
(\cite{HutererLinder06}, Fig. 4).

The survey parameters are summarised in Table \ref{Surveys},
including the photometric redshift error $\sigma_z$ and the number
of sources per square arcminute, $n_0$. The Fisher matrices for the
four experiments are available at http://www.roe.ac.uk/$\sim$afh. As
there is a degeneracy between $w_0$, $w_a$ and $\gamma$ for
\emph{Planck}+WL, better constraints on the Universe expansion
history lead to a better determination of $\gamma$ and therefore
better model selection power.  For probes of the expansion history
we consider supernovae and a sample of $2000$ supernovae type 1a at
$0<z\leq 1.8$ (see \cite{Virey} and \cite{Yeche}) as produced
by \emph{SNAP} (\cite{SNAP,SNAPWL}) or the Advanced Dark Energy
Probe Telescope (\emph{ADEPT}, a candidate for the NASA Joint Dark
Energy Mission; C. Bennett, private communication). For BAO we
consider a wide survey contemplated by WFMOS (\cite{Bassett05}) or \emph{ADEPT}.

\begin{table}
\begin{center}
\begin{tabular}{|l|c|c|c|c|}
 \hline
Survey& Area/sq deg&$z_m$&$n_0$/sq '&$\sigma_z(z)$\\
 \hline
DES&$5$,$000$&$0.80$&$10$&$0.05(1+z)$\\
 \hline
PS1&$30$,$000$&$0.75$&$5$&$0.06(1+z)$\\
 \hline
\emph{WL$_{NG}$}&$20$,$000$&$0.90$&$35$&$0.025(1+z)$\\
 \hline
\end{tabular}
\caption{The critical survey parameters for the Weak Lensing
experiments considered. DES is the Dark Energy Survey, PS1 is the
Pan-STARRS single telescope 3$\pi$ survey, and $WL_{NG}$ is a
hypothetical `next-generation' imaging survey, such as \emph{DUNE}, \emph{SNAP}, or LSST. $z_m$ is the median
redshift, $n_0$ the number of sources per square arcminute, and
$\sigma_z$ is the assumed photometric redshift error. } \label{Surveys}
\end{center}
\end{table}

From these Fisher matrices (the Fisher matrix of a combination of
independent data sets is the sum of the individual Fisher matrices),
we compute the ratio of expected evidences assuming that the true model
is a DGP braneworld, and take a prior range $\Delta\gamma=1$. Table
\ref{Results} shows the expected evidence for the 3D weak lensing
surveys with and without \emph{Planck}.

We find that $\ln B$ obtained for the standard General Relativity
model is only $\sim 1$ for DES+\emph{Planck}, whereas for
Pan-STARRS$+$\emph{Planck} we find that $\ln B\sim 2$, for
Pan-STARRS$+$\emph{Planck}$+$\emph{SN}$+$\emph{BAO} $\ln B \sim
3.61$ and for \emph{WL$_{NG}$}$+$\emph{Planck}, $\ln B$ is a
decisive $52.2$. Furthermore a \emph{WL$_{NG}$} experiment could
still decisively distinguish Dark Energy from modified gravity
without \emph{Planck}. The expected evidence in this case scales
proportionally as the total number of galaxies in the survey.
Pan-STARRS and \emph{Planck} should be able to determine the
expansion history, parametrised by $w(a)$ to very high accuracy in
the context of the standard General Relativity cosmological model,
with an accuracy of $0.03$ on $w(z\simeq 0.4)$, it will be able to
substantially distinguish between General Relativity and the
simplification of the DGP braneworld model considered here, although
this does depend on there being a strong CMB prior.

The relatively low evidence from DES+\emph{Planck} in comparison to
Pan-STARRS+ \emph{Planck} is due to the degeneracy between the
running of the spectral index $\alpha$, and $\gamma$. The larger
effect volume of Pan-STARRS, in comparison to DES, places a tighter
constraint on this degeneracy. This could be improved by on-going
high resolution CMB experiments.

\begin{figure}
  \includegraphics[width=0.45\textwidth]{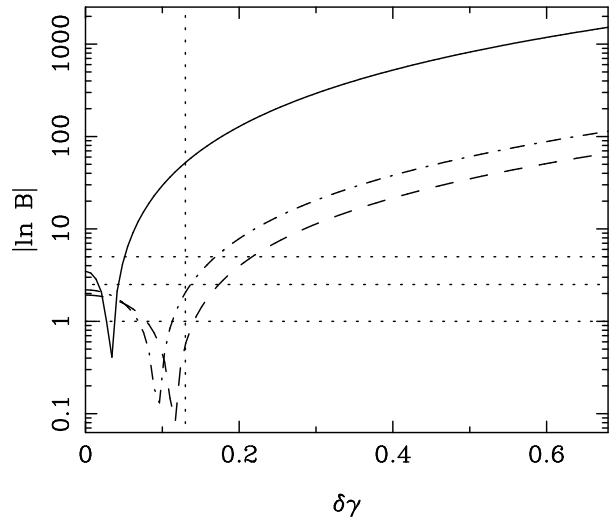}\\
  \caption{The expected value of $|\ln B|$ from WL$_{NG}$ (solid),
Pan-STARRS (dot-dashed) and DES (dashed), in combination with CMB
constraints from \emph{Planck}, as a function of the difference in
the growth rate between the modified-gravity model and General
Relativity.  The crossover at small $\delta\gamma$ occurs because
Occam's razor will favour the simpler (General Relativity) model
unless the data demand otherwise. To the left of the cusps, GR would
be likely to be preferred by the data. The dotted vertical line
shows the offset of the growth factor for the DGP model. The
horizontal lines mark the boundaries between `inconclusive',
`significant', `strong', and `decisive' in Jeffreys' (1961)
terminology.}
  \label{dgamma}
\end{figure}

Alternatively, we can ask the question of how different the growth
rate of a modified-gravity model would have to be for these
experiments to be able to distinguish the model from General
Relativity, assuming that the expansion history in the modified
gravity model is still well described by  the $w_0,w_a$
parametrization.  This is shown in Fig.\ref{dgamma}.  It shows how
the expected evidence ratio changes with progressively greater
differences from the General Relativistic growth rate.  We see that
a WL$_{NG}$ survey could even distinguish `strongly'
$\delta\gamma=0.048$, Pan-STARRS $\delta\gamma=0.137$ and DES
$\delta\gamma=0.179$.  Note that changing the prior range
$\Delta\gamma$ by a factor 10 changes the strong/decisive boundary for $\delta\gamma$ (for $WL_{NG}+$Planck) by $\sim 0.012$, so the dependence on the prior range is rather
small.

If one prefers to ask a frequentist question, then a combination of
WL$_{NG}$+{\em Planck}+BAO+SN should be able to distinguish
$\delta\gamma=0.13$, at $10.6\sigma$.  Results for other experiments
are shown in Fig. \ref{Results}.

The case for a large, space-based 3D weak lensing survey is
strengthened, as it offers the possibility of conclusively
distinguishing Dark Energy from at least some modified gravity
models.

\begin{table}
\begin{center}
\begin{tabular}{|l|c|c|c|}
 \hline
Survey&$\nu$& $|\ln B|$&\\
\hline
DES+\emph{Planck}+BAO+SN& 3.5 &$1.28$&substantial\\
DES+\emph{Planck}& 2.2 &$0.56$&inconclusive\\
DES& 0.7 &$0.54$&inconclusive\\
 \hline
PS1+\emph{Planck}+BAO+SN& 2.9 &$3.78$&strong\\
PS1+\emph{Planck}& 2.6 &$2.04$&substantial\\
PS1& 1.0 &$0.62$&inconclusive\\
 \hline
\emph{WL$_{NG}$}+\emph{Planck}+BAO+SN &10.6&$63.0$&decisive\\
\emph{WL$_{NG}$}+\emph{Planck}& 10.2 &$52.2$&decisive\\
\emph{WL$_{NG}$}& 5.4 &$11.8$&decisive\\
 \hline
\end{tabular}
\caption{The evidence ratio for the three weak lensing experiments
considered with and without \emph{Planck}, supernova and BAO priors.
$WL_{NG}$ is a next-generation space-based imaging survey such as
proposed for \emph{DUNE} or \emph{SNAP}. For
completeness, we also list the frequentist significance $\nu\sigma$
with which GR would be expected to be ruled out, if the DGP
braneworld were the correct model. } \label{Results}
\end{center}
\end{table}

\section{Conclusions}

We have shown in this paper how one can compute the Bayesian
Evidence under the assumption of a Gaussian likelihood surface,
taking account of the fact that assuming the wrong model choice can
affect the best-fitting values of parameters in the models.  The
assumption of a Gaussian likelihood is the same as used in the
Fisher matrix approach to forecasting parameter errors, and we find
that the Evidence can be calculated directly from the Fisher matrix
alone, and with very little extra computation.  An important caveat
is that the assumption that the likelihood is a multivariate
Gaussian is very strong, and deviations from this could easily
change the evidence substantially.  Nevertheless, the assumption is the same
as is used in estimating marginal errors with Fisher matrices; for the CMB part used here we find agreement with 30\% accuracy (note our Fisher matrices for Planck are conservative).  However, this method should
only be regarded as a first step, to identify which experimental
setups are worth exploring with more detailed investigations.  For the 3D weak lensing
part of the study in particular, this method is extremely valuable, as simulating the surveys and subsequent analysis is an enormously time-consuming task.

We have also shown how future observations of the Cosmic Microwave
Background, 3D weak lensing and probes of the expansion
history of the Universe offer considerable promise of distinguishing decisively between
Dark Energy models and modifications to General Relativity. In
particular, a combination of \emph{Planck} (CMB) and  proposed
space-based wide field imaging (weak lensing) surveys should be able
decisively to distinguish a Dark Energy General Relativity model
from a DGP modified-gravity model with natural log of the expected
evidence ratio $\ln B \simeq 50$.

Surveys such as \emph{DUNE/SNAP/}LSST, Pan-STARRS and DES, in
combination with Planck, should be able to distinguish `strongly'
between General Relativity and minimally-modified gravity models
with growth rates larger than 0.60, 0.69 and 0.73 respectively. The
addition of probes of the expansion history of the Universe, such as
supernova and baryon acoustic oscillation surveys help lift residual
degeneracies and thus to distinguish `strongly' models with growth
rates larger than 0.59, 0.67 and 0.71 respectively.

\section*{Acknowledgements}
We thank Andy Taylor, Adam Amara and Sarah Bridle for helpful
discussions, Viviana Acquaviva, Christophe Yeche, Filipe Abdalla, Jiayu Tang and Jochen
Weller for CMB Fisher matrix comparison, and Ramon Miquel, Hiranya
Peiris, Chuck Bennett and Pia Mukherjee for useful comments on an
early draft of the paper.  We also thank the referee, Roberto
Trotta, for a helpful report and interesting discussions. TDK acknowledges the University of
Oxford observational cosmology rolling grant. LV is supported by
NASA grant ADP03-0000-009 and NASA grant ADP04-0000-0093.

\bibliographystyle{mn2e}

\end{document}